\documentclass[letterpaper]{article} % DO NOT CHANGE THIS
\usepackage{aaai25}  % DO NOT CHANGE THIS
\usepackage{times}  % DO NOT CHANGE THIS
\usepackage{helvet}  % DO NOT CHANGE THIS
\usepackage{courier}  % DO NOT CHANGE THIS
\usepackage[hyphens]{url}  % DO NOT CHANGE THIS
\usepackage{graphicx} % DO NOT CHANGE THIS
\usepackage{natbib}  % DO NOT CHANGE THIS AND DO NOT ADD ANY OPTIONS TO IT
\usepackage{caption} % DO NOT CHANGE THIS AND DO NOT ADD ANY OPTIONS TO IT
\usepackage{amsmath}
\usepackage{amssymb}
\usepackage{multirow}
\usepackage{tabularx}
\usepackage{newfloat}
\usepackage{listings}
\usepackage{algorithm}
\usepackage{algpseudocode}
\DeclareCaptionStyle{ruled}{labelfont=normalfont,labelsep=colon,strut=off} % DO NOT CHANGE THIS
\floatstyle{ruled}
\newfloat{listing}{tb}{lst}{}
\floatname{listing}{Listing}
\title{Secure Multiparty Generative AI}
\author{
    Manil Shrestha\textsuperscript{\rm 1,2}\thanks{Work performed during employment at Block Entropy Inc}, Yashodha Ravichandran\textsuperscript{\rm 1,2*}, Edward Kim\textsuperscript{\rm 1,2}
}
\affiliations{
    \textsuperscript{\rm 1}Entropy Labs, Block Entropy, Pennsylvania, USA\\
    \textsuperscript{\rm 2}Drexel University, Department of Computer Science, Pennsylvania, USA
    ms5267@drexel.edu, yr82@drexel.edu, ek826@drexel.edu
}

\usepackage{bibentry}
\begin{document}
\maketitle
\begin{abstract}
The field of generative artificial intelligence has recently had an unprecedented impact and adoption. These generative machine learning models can generate data similar to the data they were trained on with remarkable fidelity.  Common generative tasks include creating realistic images from text prompts via stable diffusion and generating text via large language models.  However, as usage of these tools skyrockets, the amount of sensitive information being exposed to these models and centralized model providers is alarming.  For example, confidential source code from Samsung suffered a data leak as the text prompt to ChatGPT encountered data leakage.  An increasing number of companies are restricting the use of LLMs (Apple, Verizon, JPMorgan Chase, etc.) due to data leakage or confidentiality issues.  Also, an increasing number of centralized generative model providers are restricting, filtering, aligning, or censoring what can be used.  Midjourney and RunwayML, two of the major image generation platforms, restrict the prompts to their system via prompt filtering.  Certain political figures are restricted from image generation, as well as words associated with women's health care, rights, and abortion.

In our research, we present a secure and private methodology for generative artificial intelligence that does not expose sensitive data or models to third-party AI providers.  Our work modifies the key building block of modern generative AI algorithms, e.g. the transformer, and introduces confidential and verifiable multiparty computations in a decentralized network to maintain the 1) privacy of the user input and obfuscation to the output of the model, and 2) introduce privacy to the model itself.  Additionally, the sharding process reduces the computational burden on any one node, enabling the distribution of resources of large generative AI processes across multiple, smaller nodes.  We show that as long there \textit{exists one honest node} in the decentralized computation, security is maintained.  We also show that the inference process will still succeed if only a majority of the nodes in the computation are successful.  Thus, our method offers both secure and verifiable computation in a decentralized network.
\end{abstract}


\section{Introduction}
\begin{figure}
    \centering
    \includegraphics[width=0.9\linewidth]{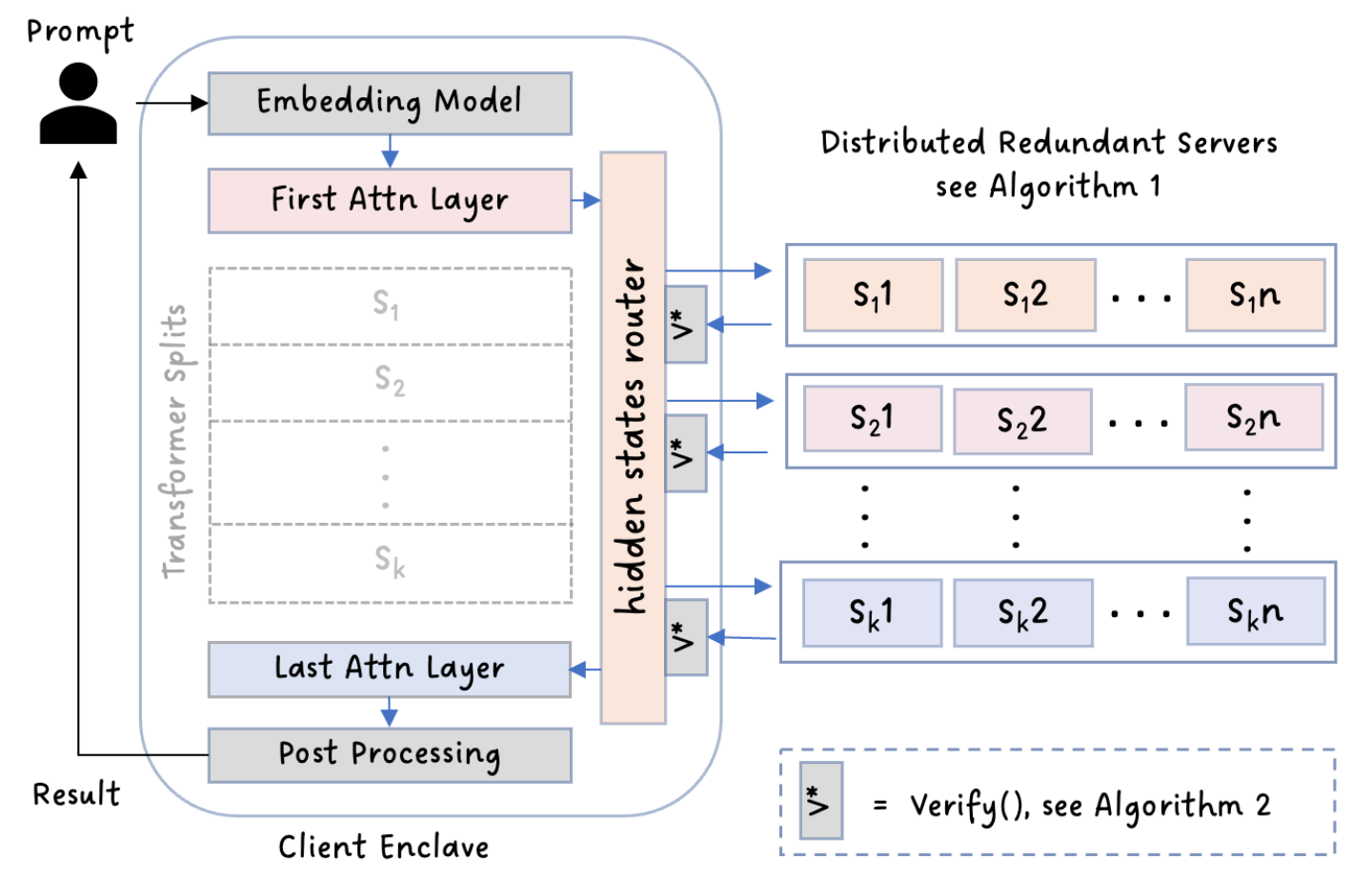}
    \caption{Illustration of the Secure Multi-Party Computation (SMPC) architecture for a transformer-based generative AI model. Embedding model along with first, and last attention layers of transformer is securely hosted within the client enclave. Remaining layers are divided into \textit{k} `splits'.  Each of the \textit{k} splits is distributed across decentralized servers, where one or more servers may host the same split.}
    \label{fig:front_page_graphics}
    \vspace{-0.4cm} % Reduces the vertical space after the figure
\end{figure}

The explosive growth of generative artificial intelligence technologies has introduced unprecedented adoption of AI-related use cases. Generative models are being used to create images, music, text, code, virtual environments, speech, and more \cite{rombach2022high, hadjeres2017deepbach, achiam2023gpt, li2022competition, ling2015deep, tassa2018deepmind}.  However, the widespread use has also raised significant privacy concerns. The core of the problem is that the data being input into these systems for inference and training may contain confidential, sensitive, or explicit material that should be private. Further, training a generative model requires enormous amounts of data, often collected from users who may or may not be fully aware of how their information will be used.

In our research, we investigate a Secure Multi-Party Computation (SMPC) method to protect user privacy in the generative AI inference process, see Figure \ref{fig:front_page_graphics}.  SMPC is a subfield of cryptography that allows multiple parties to jointly compute a function over their inputs while keeping these inputs private \cite{yao1982protocols, goldreich1998secure,wigderson1988completeness}. These mechanisms allow a group of people, machines, or organizations to collaborate on a calculation without revealing their individual data to each other. We believe there are data-critical situations where sharing data openly would compromise privacy or security.

The core idea behind SMPC is to distribute computations in such a way that no single party gains access to more information than they are supposed to know. This can be achieved through algorithmic, mathematical, or cryptographic techniques that ``split'' the input into shares that are then distributed among the participating parties. The computation is carried out on these shares, and the final result is reconstructed only when necessary, ensuring that individual inputs remain confidential. 

Our contributions are three-fold.  First, we propose an algorithm to optimally shard a transformer-based generative AI model that hides the input, obfuscates the output, and protects the intellectual property of the model used in a decentralized network. 
 Second, we present results on an SMPC distributed inference network that demonstrate the tradeoff that exists between privacy and performance, and third, we propose a novel verification algorithm that can ensure with high confidence that the nodes participating in the SMPC network are performing as expected.  \textit{It is important to note that there are several assumptions and conditions for our contributions to hold.}  For privacy of input and obfuscation of the output, the assumption is that the computational nodes do not know, and do not have access to the embedding model being used.  The second assumption is that there is at least one honest node in the SMPC computational chain in order to protect the IP of the model.  And third, there needs to be a majority of honest nodes in the computational chain for the verification algorithm to reach the correct consensus.  In the methodology and experiments, we will elucidate the strengths and limitations of our approach.  

\section{Background}
In the area of AI privacy and verification, there are many alternative approaches available, including homomorphic encryption, zk-proofs, and confidential computing.  Each one of these fields has a large body of literature, and will be briefly summarized in this section; while these methods have strong guarantees, they are all currently intractable solutions for a deployed privacy-preserving AI today.  \

\subsection{Related Work in Privacy Preserving AI}
\textit{Homomorphic Encryption -} Homomorphic encryption is a class of methods that allow one to perform operations on information, while it is still encrypted.  There are different levels of homomorphic encryption including Partially Homomorphic Encryption (PHE), of single operations (like addition or multiplication), and Fully Homomorphic Encryption (FHE), enabling arbitrary computations in the encrypted space.  While this is ideal, fully homomorphic encryption schemes suffer from high computational complexity, low efficiencies, and inadequacy of deployment in real-world scenarios, making them impractical for real-time applications \cite{marcolla2022survey}. \\
\textit{ZK-Proofs - } A zero-knowledge proof is a method by which one party (the prover) can prove to another party (the verifier) that they know a value x, without conveying any information apart from the fact that they know the value x \cite{goldwasser2019knowledge}. Zero-knowledge proofs must satisfy three key properties of completeness (if the statement is true, an honest verifier will be convinced of this fact by an honest prover), soundness (if the statement is false, no cheating prover can convince an honest verifier that it is true, except with some small probability), and zero-knowledge, (if the statement is true, the verifier learns nothing other than the fact that the statement is true).   While there have been significant strides in reducing the proving time of zero knowledge, i.e. one scheme reduces the proving time approximately 18,000x for the canonical CNN models on VGG16 \cite{lee2024vcnn}, the reality is that the zk-snark (succinct argument of knowledge) still takes 44 hours to complete, rather than 10 years.  Several orders of magnitude in performance are still needed for a practical and deployable solution. \\
\textit{Confidential Computing -} Confidential computing focuses on protecting data while it's in use, complementing existing security measures that protect data at rest and in transit. This technology creates a secure enclave within a computer's processor, isolating sensitive data and code from the rest of the system. By doing so, it shields the data from unauthorized access, even in the event of a system compromise \cite{mulligan2021confidential}.  Confidential computing ensures that even cloud service providers cannot access the data being processed on their systems by employing hardware-based trusted execution environments (TEEs). These TEEs, also known as secure enclaves, provide an isolated region of memory where code can be executed and data can be processed securely. The contents of this secure enclave are encrypted and can only be decrypted within the CPU itself, making it extremely difficult for malicious actors to access or tamper with the data.  Unfortunately, support for these TEEs is not widespread and has little to no support for GPU processing in consumer hardware.  Only recently with the H100 NVidia GPU can one perform tasks in a secure enclave.\\
\textit{Multiparty Computation (MPC) in ML -} 
There has been significant progress in making ML inferences more secure and private. Initially, much of the work focused on ML algorithms such as regression and CNN-based vision models and used arithmetic secret sharing methods \cite{ghodsi2021circa, kumar2020cryptflow, wagh2020falcon, patra2020blaze, knott2021crypten}. Later works highlighted the performance issues of MPCs when applied to transformer-based models \cite{ wang2022characterization}. As a result, efforts have been directed toward improving the performance of existing MPC frameworks \cite{li2022mpcformer,cho2022selective}. All of the MPC frameworks mentioned here focus on distributing secret `shares' of input to different parties holding `shares' of a model.

\subsection{Related Work in Distributed Computing}
As for the distribution of work across multiple nodes, a common practice is the sharding of a model.
Sharding a model is an industry standard when a model needs to be split across multiple devices, either due to the model's size exceeding the capacity of a single machine or to improve the system's scalability \cite{dean2012large, rajbhandari2020zero, lepikhin2020gshard}. Several open-source libraries exist for model sharding, however, the key focus of these methods is to to improve model performance and scalability for training and inference. Some of them are discussed briefly in this section. DeepSpeed is an open-source deep learning library developed by Microsoft that powers unprecedented scale and speed for training, inference, and model compression using techniques such as Zero Redundancy Optimizer (ZeRO) \cite{rajbhandari2020zero}. DeepSpeed-Inference uses model parallelism, high-performance kernels, and memory optimization to enable low-latency inference. Any pre-trained model can be loaded with a user-defined parallelism parameter. DeepSpeed will automatically shard the model across multiple GPUs for inference. \cite{aminabadi2022deepspeed}. Fully Sharded Data Parallel (FSDP) is a training technique in the PyTorch distributed module, designed for models that cannot fit into a single GPU's memory. Inspired by \cite{xu2020automatic} and ZeRO Stage 3, FSDP splits model parameters, gradients, and optimizer states across multiple GPUs. It uses an all-gather operation to collect parameters and gradients for processing and a reduce-scatter operation to shard gradients across GPUs after the backward pass from Nvidia's NCCL library \cite{nvidiaNVIDIACollective} \cite{zhao2023pytorch}. %Text Generation Inference (TGI) is a framework from Hugging Face to deploy and LLMs. TGI provides efficient text generation capabilities for the most open-source LLMs. It consists of a router that receives the client requests, buffers them, creates batches, and prepares gRPC calls to a model server that then processes inference across all the model shards \cite{huggingfaceTextGeneration}.
  However, none of these libraries and methods tackle the core problem of privacy and verification of work in a trustless manner.  Thus, we propose a Secure Multiparty Computation approach to distributed inference.

\section{Methodology}
%As discussed in the previous section, there have been numerous works on model sharding; however, these often lack focus on the security and privacy of the model and the user's inputs in a decentralized, distributed scenario. 
In this section, we propose a Secure Multi-Party Computation (SMPC) architecture within the context of a transformer-based generative AI model. We demonstrate that given a proprietary model, both the model and user prompts remain secure in a distributed setting.

% Custom Comment to indicate parallel processes
% Add these lines to define parallel for loop
\algnewcommand\algorithmicparfor{\textbf{parallel for}}
\algdef{SE}[PARFOR]{ParFor}{EndParFor}[1]{\algorithmicparfor\space#1\space\algorithmicdo}{\algorithmicend\ \algorithmicparfor}

\begin{algorithm}[!t]

\caption{SMPC in decentralized redundant servers}
\label{alg:smpc_algorithm}
\begin{algorithmic}[1]
        \Statex \textbf{Input:} User Prompt ($\mathcal{P}_i$)
        \Statex \textbf{Output:} Model Output ($\Omega_i$) 
        \State $\mathcal{T}_i \gets \mathcal{E}(\mathcal{P}_i)$  \algorithmiccomment{$\mathcal{E}$ is the embedding model}
        \State $\mathcal{H}_i \gets \Phi_0(\mathcal{T}_i)$ \algorithmiccomment{$\Phi_0$ is the first attn layer}
        \For {$j = 1$ \textbf{to} $k$} \Comment{$k$ is the no. of distributed splits}
        \State $X \gets [\ ]$ \Comment{Initialize $X$ as an empty list}
        \ParFor{$r = 1$ \textbf{to} $n$}
            \State $X \gets X.append(\Phi_{j}^{r}(\mathcal{H}_i))$
        \EndParFor
        \State $\mathcal{H}_i \gets Verify(X)$ \Comment{See  Algorithm~\ref{alg:verify_algorithm}}
    \EndFor
    \State $\Psi_i \gets \Phi_{l}(\mathcal{H}_i)$ \algorithmiccomment{$\Phi_l$ is the last attn layer}
    \State $\Omega_i \gets \xi(\Psi_i)$ \algorithmiccomment{$\xi$ denotes post processing}
    \State \textbf{return} $\Omega_i$
    \end{algorithmic}
\end{algorithm}
\begin{algorithm}[!t]
\caption{$Verify$ (Redundant servers consensus)}
\label{alg:verify_algorithm}
\begin{algorithmic}[1]
\Statex \textbf{Input:} Results from redundant servers $X$
\Statex \textbf{Output:} Verified result $\mathcal{H}_i \in X$
\State $\Bar{X}, Z \gets [\ ], [\ ]$ \Comment{Initialize X and Z as empty lists}
\For{$x \in X$}
    \State $Z \gets Z.append(\phi(x))$ \Comment{$\phi$ is a hashing function}
\EndFor
\For{$h \in Z$}
    \State $\textit{countAgree} \gets 0$
    \For{$j \in Z, j \neq h$}
        \State $\textit{sim} \gets \alpha(h,j)$ \Comment{$\alpha$ is a similarity function}
        \If{$\textit{sim} > \textit{threshold}$}
            \State $\textit{countAgree} \gets \textit{countAgree} + 1$
        \EndIf
    \EndFor
    \If{$\textit{countAgree} > \lfloor \frac{|X|}{2} \rfloor$}
        \State $\bar{X} \gets \bar{X}.append(x_h)$ \Comment{$x_h$ has hash $h$}
    \EndIf
\EndFor
\If{$| \Bar{X}| > \lfloor \frac{|X|}2 \rfloor$} 
    \State $\mathcal{H}_i \gets mode(\Bar{X})$
    \State \Return $\mathcal{H}_i$
\Else
    \State \textbf{raise} Exception(``Faulty computation")
\EndIf

\end{algorithmic}
\end{algorithm}
\newcolumntype{P}[1]{>{\centering\arraybackslash}p{#1}}

\begin{table*}[htbp]
\scriptsize
\centering
\begin{tabular}{|P{0.3cm}|P{0.8cm}|P{1.3cm}|P{1.32cm}|P{1.3cm}|P{1.32cm}|P{1.2cm}|P{1.3cm}|P{1.2cm}|P{1.5cm}|P{1.2cm}|}
\hline
&\textbf{Splits ($k$)} & \textbf{ VRAM\newline(Client)}&\textbf{Avg. VRAM\newline(per server)\newline (MB)}& \textbf{ Params\newline(client)} &\textbf{Avg. Params\newline(server)} & \textbf{Image Size /Token Length} & \textbf{Total \newline Time (s)} & \textbf{Time/Iter.(s) \newline /\newline Token/s} & \textbf{Network BW per Iter.\newline (MB)} & \textbf{Verification Time (s)} \\ \hline

\multirow{9}{*}{\rotatebox{90}{Stable Diffusion 3 Medium}}& \multirow{3}{*}{0} & \multirow{ 3}{*}{6238} & \multirow{3}{*}{-}& \multirow{ 3}{*}{2.930  B} & \multirow{3}{*}{-} & $256^2$ & 0.99 & 0.03 & - & - \\
\cline{7-11}
&& & & & & $512^2$ & 1.86 & 0.06 & - & - \\ 
\cline{7-11}
&& & & & & $1024^2$ & 6.92 & 0.23 & - & - \\

\cline{2-11}

& \multirow{3}{*}{1} & \multirow{ 3}{*}{2648} & \multirow{3}{*}{3964}& \multirow{ 3}{*}{1.061 B} & \multirow{3}{*}{1.869 B} & $256^2$ & 4.50  & 0.16 & 4.82& 0.92 \\
\cline{7-11}
&& & & & & $512^2$ & 7.34 & 0.26 & 13.81& 1.09\\ 
\cline{7-11}
&& & & & & $1024^2$ & 29.02 & 1.02 & 49.82 & 4.99\\

\cline{2-11}

& \multirow{3}{*}{2} & \multirow{ 3}{*}{2936} & \multirow{3}{*}{2116}& \multirow{ 3}{*}{1.061 B} & \multirow{3}{*}{0.935 B} & $256^2$ & 7.52 & 0.26& 9.64& 1.78\\
\cline{7-11}
& & & & & & $512^2$ & 13.15 & 0.46 & 27.62 &  2.21\\ 
\cline{7-11}
& & & & & & $1024^2$ & 49.99 & 1.77 & 99.64& 9.49\\
\hline
\cline{1-11}
\cline{1-11}

\multirow{9}{*}{\rotatebox{90}{Llama 3.1 8B}}&\multirow{3}{*}{0} & \multirow{ 3}{*}{16474} & \multirow{3}{*}{-}& \multirow{ 3}{*}{8.030 B} & \multirow{3}{*}{-} & 50 & 1.77 & 38 & - & - \\
\cline{7-11}
& & & & & & 100 & 3.29 & 40 & - & - \\ 
\cline{7-11}
& & & & & & 150 & 4.96 & 40 & - & - \\
\cline{2-11}

&\multirow{3}{*}{1} & \multirow{ 3}{*}{3262} & \multirow{3}{*}{13538}& \multirow{ 3}{*}{1.486 B} & \multirow{3}{*}{6.543 B} & 50 & 12.07 & 4.34 & 7.75& 0.14 \\
\cline{7-11}
&& & & & & 100 & 28.89 & 3.57 & 14.31 & 0.30 \\ 
\cline{7-11}
&& & & & & 150 & 57.18 & 2.70 & 20.86 & 0.46 \\

\cline{2-11}

&\multirow{3}{*}{2} & \multirow{ 3}{*}{3262} & \multirow{3}{*}{6932} &\multirow{ 3}{*}{1.486 B} & \multirow{3}{*}{3.272 B} & 50 & 43.99 & 1.15 & 15.50& 0.60 \\
\cline{7-11}
&& & & & & 100 & 122.13 & 0.83 & 28.62 & 1.22 \\ 
\cline{7-11}
&& & & & & 150 & 212.01 & 0.714 & 40.07 & 1.77 \\
\hline

\end{tabular}
\caption{Performance of the proposed SMPC architecture on the Stable Diffusion 3 Medium (\textit{top}) and Llama3.1 8B model (\textit{bottom}). Setting with $k=0$ is the base case with the model on a single machine. $k=1$ is when the client hosts the embedding models, first and last of the attention layers, while all other attention layers are hosted on a single server. This enables hiding the prompt from the third-party server. Setting $k=2$ extends the $k=1$ case by splitting the layers between two servers, securing privacy to both the prompt and model weights. For splits $k=1$ and $k=2$, the number of independent verifiers ($n$) is set to 3.}
\label{tbl:results_smpc}
\end{table*}

\subsection{System Architecture}
We propose a single-client, multi-server architecture. The client refers to a system environment controlled by the model owner, which could be within the organization's network or hosted in a secure environment. Servers represent distributed nodes that host small parts of the model. In the case of transformer-based models, each server hosts one or more attention layers. The model owner provides each server node with its respective model, without revealing which specific layer(s) it is computing. As long as at least one server remains honest, there is no possibility of collusion to reconstruct the entire model, ensuring the security of both the model weights and architecture. 
\subsubsection{SMPC in decentralized servers (Algorithm \ref{alg:smpc_algorithm}) -}
The prompt to a generative model, $\mathcal{P}_i$, is kept private by processing the input within the client enclave, i.e. projecting the token ids through an embedding model, $\mathcal{E}$, and transforming the vectors by the first attention layer, $\Phi_0$. Previous research has demonstrated that the projected textual embeddings alone are not entirely private and are vulnerable to adversarial attacks \cite{morris2023text}. An attacker would need access to the prompt and the corresponding embeddings to execute a black-box attack \cite{papernot2017practical}. However, in our design, only authorized users within the organization have permission to prompt the system, and the data leaving the client enclave for the external network consists of hidden states that are further transformed by the first attention layer.

\begin{figure}[h]
    \centering
    \includegraphics[width=0.9\linewidth]{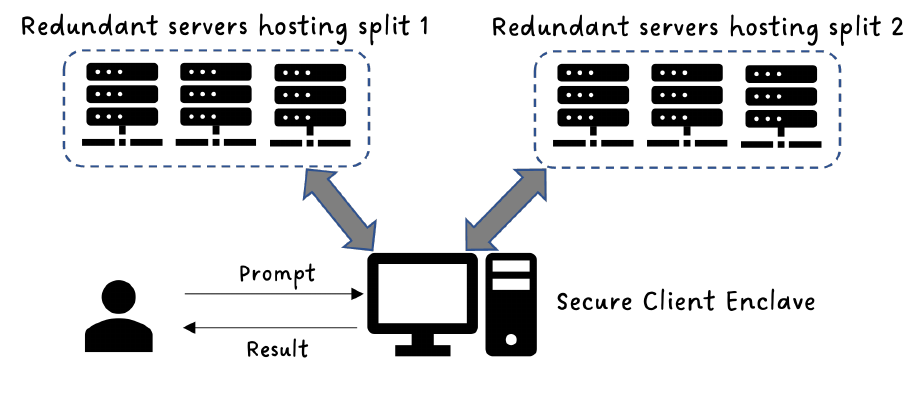}
    \caption{Illustration of SMPC architecture flow with number of splits $k=2$ and redundant nodes $n=3$.}
    \label{fig:star_architecture}
    \vspace{-0.5cm} % Reduces the vertical space after the figure
\end{figure}

A model divided into $k$ splits contains a variable number of attention layers. The client orchestrates the processing order, as it alone knows which servers host specific layers. The servers do not communicate directly with each other. This is a design choice essential to maintain ambiguity about the model's distribution across servers. After the servers complete the calculations for the final split, the client handles the remaining processing, including the final attention layer $\Phi_l$ and any post-processing $\xi$, see Figure \ref{fig:star_architecture}.

\subsubsection{Verification of work (Algorithm \ref{alg:verify_algorithm}) -}

When distributing work to a third party, there is always a risk of dishonesty. Verification entails providing a convincing argument that the system will behave correctly across different scenarios \cite{xiang2018verification}. To ensure that each split $j$'s calculation is correct, we add redundancy of work into our design. Consensus between independent verifiers assumes that the majority of the server nodes are honest. There are $n$ servers with the same attention layers performing identical computations in parallel. Server $r$ with split $j$ has a model $\Phi_{j}^{r}$. As the client receives results from $n$ servers hosting split $j$, a verification process ensures that the results are valid. The outputs from each of the $n$ servers are first hashed using a locally sensitive hashing (LSH). This results in $n$ sets of hashes, and for each hash $h$, we compare the Hamming distance-based similarity to the other $n-1$ hashes. 

Let $h_1$ and $h_2$ be two hash values of the same length $l$. The hash similarity $sim(h_1, h_2)$ is defined as:
\begin{equation}
sim(h_1, h_2) = 1 - \frac{1}{l}\sum_{i=1}^l \delta(h_{1,i}, h_{2,i})
\end{equation}

\begin{equation}
\delta(x, y) = \begin{cases} 
1 & x \neq y \\ 
0 & x = y 
\end{cases}
\end{equation}

The decision to use an LSH was made due to structural uniformity within the outputs from attention layers. If the similarity score exceeds a predefined threshold for more than $\lfloor \frac{n}{2} \rfloor$ other redundant servers, we consider the work correct and add it to a list $\Bar{X}$ of valid results with similarity within the threshold.

To ensure consensus, we require $|\Bar{X}| > \frac{|X|}{2}$; otherwise, we deem the computation untrustworthy and raise an exception. If this threshold is met, we select the mode of $\Bar{X}$ as the output for split $j$, or arbitrarily choose any result if all values in $\Bar{X}$ are unique.

% TODO: Discussion on the scalability
% TODO: How is it different from just basic sharding. Need to mention it.

\section{Experiments}
For our experiments, we set up a heterogeneous, decentralized GPU network. The client, running on an NVIDIA A40 GPU, is connected to the servers via the internet through designated network ports. The servers, that host the splits, consist of NVIDIA A40, GeForce RTX 3080 Ti, and GeForce RTX 4090 GPUs. 

We conducted experiments on two modalities of generative AI: image and text. The experimental setup and a detailed discussion of the results are provided in the following subsections.

\subsection{Experiment Setup}
For image generation, we used the Stable Diffusion 3 (SD3) Medium model to demonstrate the proposed architecture \cite{esser2024scaling}. SD3 generates detailed images from text descriptions using a latent diffusion model. The model employs 24 transformer attention layers to remove successive Gaussian noise through a sequence of denoising steps. We did not include the optional T5 text encoder in the diffusion pipeline for these experiments. Throughout the experiments, we consistently set the total number of inference steps to 28 and used 20 different prompts to run the diffusion pipeline 10 times for each setting variation to get average runtimes.

For language generation experiments with the proposed SMPC architecture, we use the Llama 3.1 8B model \cite{dubey2024llama}, which is an auto-regressive decoder-only model. It consists of 32 transformer decoder layers that work in succession to predict the next token. We generated the text for 5 different prompts for each setting to get average runtimes.
% Should we release the prompts used in supplementary material?

Table~\ref{tbl:results_smpc} shows the impact on the performance of varying the number of splits $k$ and image sizes/token length. The first setting, $k=0$, is the vanilla setup with the models running on a single client machine. In the second setting, $k=1$, for SD3, the client hosts the text encoders (CLIP-G/14 and CLIP-L/14), VAE, and the first and last attention layers, while the remaining layers are handled by third-party servers. For Llama3.1, the client hosts the tokenizer, embedding layers, and the first and last decoder layers. This configuration secures user prompts but exposes most of the model to the servers. The third setting, $k=2$, addresses this issue by not providing all the remaining attention layers to a single server but first splitting them into two distinct models, which are then anonymously distributed across multiple servers. Redundant work is implemented to ensure correctness. The number of independent verifiers is set to $n=3$ for all experiments where $k>0$. For the verification of work, hashing function $\phi$ is set to be perceptual hash (pHash) \cite{kozat2004robust} for image generation and difference hash (dHash) \cite{krawetz2013} for  text generation (see Alg. \ref{alg:verify_algorithm}).

\subsection{Tolerance in Reproducibility} 
\begin{figure}[ht!]
    \centering
    \includegraphics[width=\linewidth]{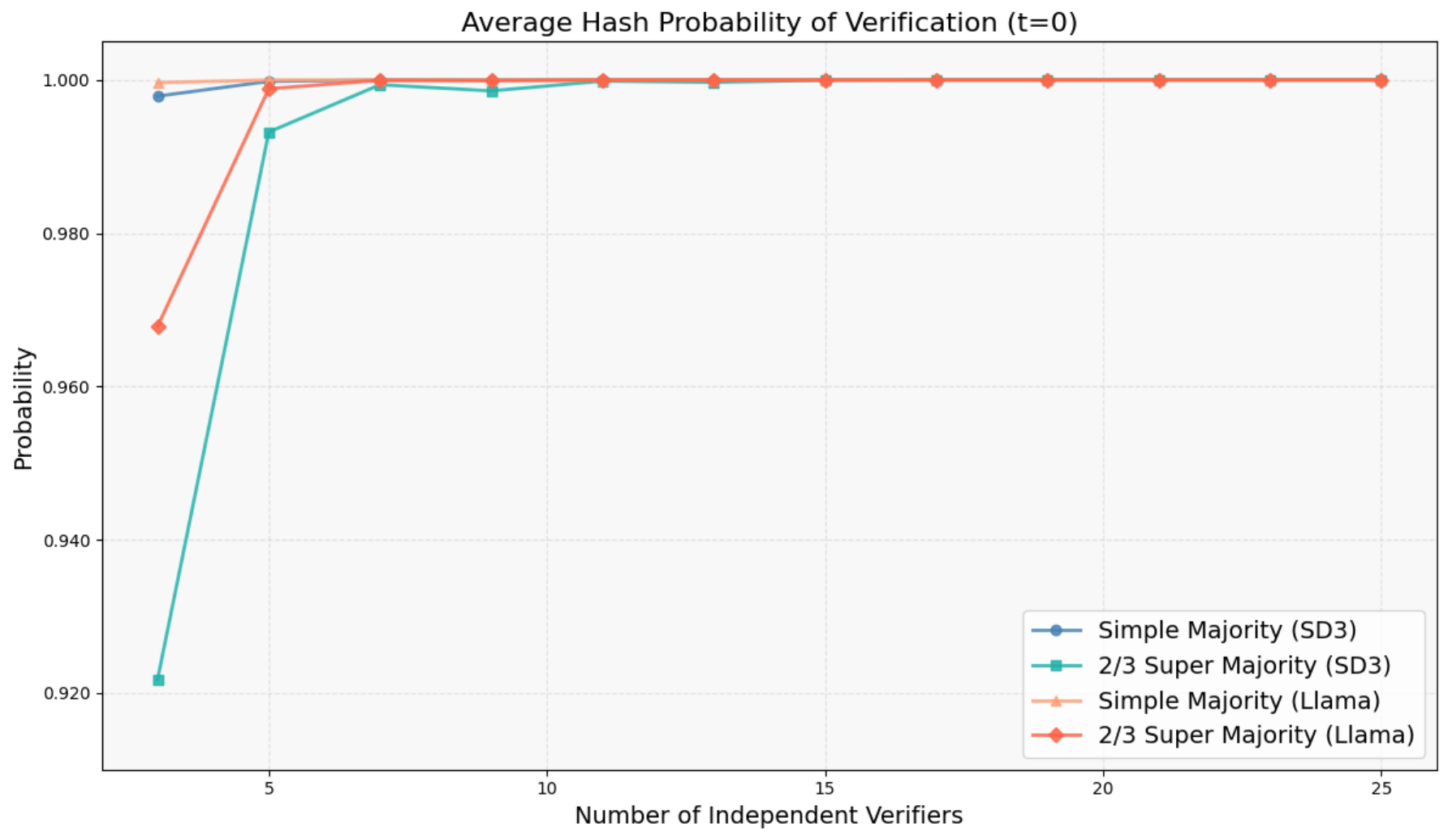}
    \caption{Graph showing the probabilities of independent verifiers detecting incorrect or fraudulent behavior under different likelihoods of deterministic generation. In the simple majority case, the majority of nodes are assumed honest, while in the super-majority case, over two-thirds of the nodes are assumed honest.}
    \label{fig:average_hash_prob}
    \vspace{-0.2cm} % Reduces the vertical space after the figure
\end{figure}

Even when two machines host the same model and have identical inputs, deviations can occur in CUDA-based computations due to inherent non-deterministic behavior \cite{PyTorchReproducibility}. In our experiments, we noticed that the LSH did not always match between the honest independent verifiers. 
We present posterior probabilities for detecting incorrect or fraudulent behavior, assuming either a majority or super-majority of nodes in the network are honest (Figure \ref{fig:average_hash_prob}).

We use, $P(X \geq k) = \sum_{i=k}^n \binom{n}{i} p^i (1-p)^{n-i}$, where the binomial distribution gives the probability of getting more than or equal to $k$ successes over $n$ independent verifiers \cite{kim2023generative}. 
In the case of image generation, we observed a matching probability of $p=97.32\%$ with a tolerance of $t=0$ (hashes matched exactly between servers). Based on this, we determined that simple majority verification achieves accuracies of 99.789\%, 99.815\%, and 99.998\% using 3, 5, and 7 independent verifiers, respectively. Supermajority verification (with $>$2/3 agreement) yields accuracies of 99.320\%, 99.938\%, and 99.857\% using 5, 7, and 9 independent verifiers, respectively.

For text generation, the matching probability was $p=98.92\%$ with $t=0$. Simple majority verification achieves accuracies of 99.965\%, 99.999\%, and 99.999\% using 3, 5, and 7 independent verifiers, respectively. Supermajority verification (with $>$2/3 agreement) yields accuracies of 96.785\%, 99.938\%, and 99.857\% using 3, 5, and 7 independent verifiers, respectively.
These results demonstrate high verification accuracy with minimal redundant work across both majority and super-majority schemes, see Figure \ref{fig:average_hash_prob}.

\subsection{Discussion}
The data in table~\ref{tbl:results_smpc} shows that as the number of splits $k$ increases, performance declines. As expected, the major bottleneck occurs in network transfer, with the number of data transfers between the client and servers increasing linearly by $2k$ with respect to the number of splits $k$. The data must be offloaded to the CPU before being sent over the internet and then reloaded into VRAM at each step, which also has a significant impact on performance. The parallelization of verifiers was possible because they all perform the same calculations. However, the servers hosting different splits must communicate sequentially, as the input for server $j$ depends on the output of server $j-1$. This limitation of the algorithm prevents a fully parallel distribution of work.

In the case of image generation, we observe that image size directly affects the processing time. This is because, in stable diffusion, the size of latent hidden states being passed around is dependent on the image size. Similarly, in text generation, the time required to generate the next token increases as the sequence lengthens. In the current setting, each server needs a new updated KV cache to be transferred over the network after the previous iteration. This is something we are currently working to optimize.

Although verification with hashes takes much less time with $n=3$, it exhibits a quadratic growth pattern in relation to the number of verifiers ($n$). In the previous section, we demonstrated that when the majority of nodes in the network are honest, a relatively small set of verifiers is sufficient to ensure reliability.

\begin{figure}[ht]
    \centering
    \includegraphics[width=1\linewidth]{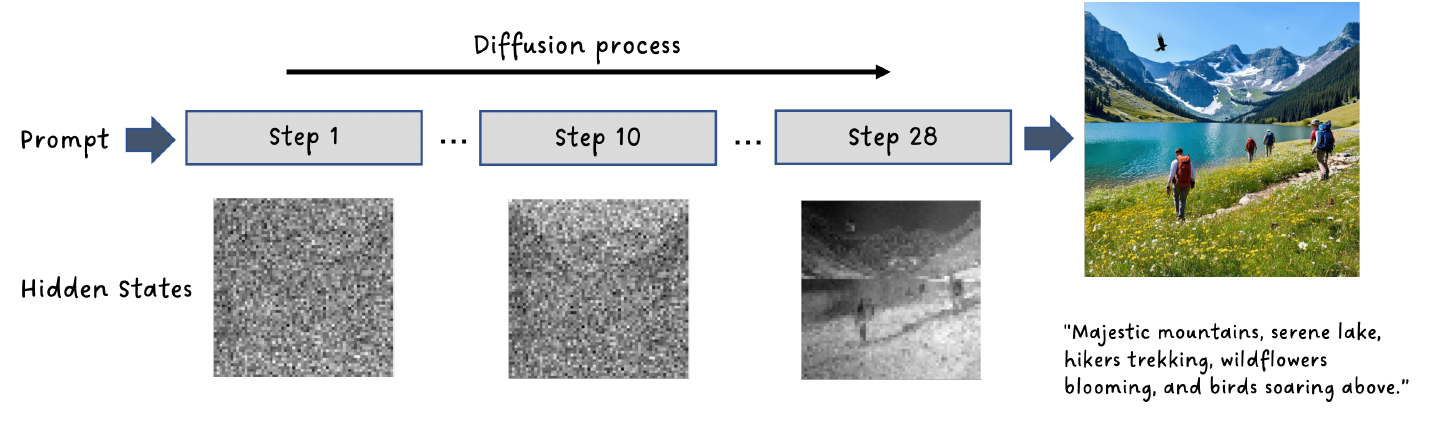}
    \caption{Visualizing the hidden states received by the client, we observe mostly noise initially, with patterns emerging in later, though still obfuscated, steps.}
    \vspace{-0.3cm} % Reduces the vertical space after the figure
    \label{fig:hidden_states}
\end{figure}

\subsection{Path to Deployment}
We are developing secure multiparty computation methods to protect the privacy of user information in our cloud platform hosted at blockentropy.ai.  On this platform, we offer API access to generative language, image, and video services.  This work is an important first step towards privacy-preserving deployment.  AI will be entering an ``HTTPS''-like revolution in the next decade where data and processing will eventually need to be protected.  Currently, all data must be processed in ``clear text'', where the computing environment can see all the inputs, models, and outputs of the system.  SMPC with secure enclaves for the client can provide privacy to the user prompts, and allow the model performing inference to be hosted in a trustless, decentralized fashion while ensuring the protection of intellectual property.  While there are trade-offs in speed and computation off-load, we are continually improving the latency of these approaches.

\section{Limitations}
Firstly, the current approach faces scalability challenges. As discussed earlier, significant latency arises from multiple data transfers over the network, which impacts performance. Additionally, in the context of image generation, the hidden states computed by the servers are not entirely concealed. As illustrated in Figure~\ref{fig:hidden_states}, visualizing the hidden states during later inference steps reveals an obfuscated yet discernible output. This indicates that while the input prompt remains secure, the generated output is not fully private. Conversely, we observed that hidden states in Llama models are not easily decipherable into meaningful information, suggesting a difference in privacy levels between modalities.

Beyond scalability and privacy, other limitations include the computational overhead associated with redundant verification processes, particularly as the number of verifiers increases. Additionally, the current method relies heavily on the assumption that at least one node is honest, which may not always hold in highly adversarial environments. Furthermore, the sequential nature of processing across server splits limits the potential for parallelism, constraining overall system efficiency.

\section{Conclusion}
We propose a Secure Multi-Party Computation (SMPC) architecture for transformer-based generative AI models. This ensures user input privacy and model intellectual property protection by securely sharding the model across multiple servers in a decentralized network. Our verification algorithm mitigates the risk of dishonest computations by leveraging redundant work and hash-based verification, achieving high accuracy with a small number of verifiers. This demonstrates that secure and verifiable computation is possible even in decentralized environments without assumed trust.
The approach has limitations in scalability and performance, particularly as the number of splits increases, causing more network latency. Future work will focus on optimizing these aspects, exploring efficient communication protocols, and ways to optimize data load/offload between devices.
Our work represents a significant step toward privacy-preserving generative AI, offering a promising path for deploying AI services securely and in a decentralized manner. As demand for AI solutions grows, protecting sensitive data and intellectual property will be increasingly critical.

\bibliography{aaai25}

\end{document}